\documentclass[lettersize,journal]{IEEEtran}
\IEEEoverridecommandlockouts
\usepackage{cite}
\usepackage{amsmath,amssymb,amsfonts,stackrel,eqnarray,}
\usepackage{algorithmic}
\usepackage{graphicx}
\usepackage{textcomp}
\usepackage{xcolor}
\usepackage{multirow}
\usepackage{steinmetz}
\usepackage{amssymb}
\usepackage{circuitikz}
\usepackage{tikz,subfigure}
\usepackage[caption=false,font=normalsize,labelfont=sf,textfont=sf]{subfig}

\DeclareMathOperator{\sinc}{sinc}

\def\BibTeX{{\rm B\kern-.05em{\sc i\kern-.025em b}\kern-.08em
    T\kern-.1667em\lower.7ex\hbox{E}\kern-.125emX}}
\usetikzlibrary{positioning}
\begin{document}

\title{Ambient FSK Backscatter Communications using LTE Cell Specific Reference Signals}

\author{Jingyi Liao,
    Xiyu Wang,
    Kalle Ruttik,
    Riku J\"{a}ntti,
    and Phan-Huy Dinh-Thuy
    \thanks {
        J. Liao, X. Wang, K. Ruttik, and R. J\"{a}ntti are with
        Department of Information and Communications Engineering,
        Aalto University,
        02150 Espoo, Finland
        (email: \{firstname.lastname\}@aalto.fi).}
    \thanks{
        Phan-Huy Dinh-Thuy is with
        Networks, Orange Innovation, Chatillon, France
        (email: dinhthuy.phanhuy@orange.com).}
    }

\maketitle

\begin{abstract}
Long Term Evolution (LTE) signal is ubiquitously present in electromagnetic (EM) background environment, which make it an attractive signal source for the ambient backscatter communications (AmBC).
In this paper, we propose a system, in which a backscatter device (BD) introduces artificial Doppler shift to the channel which is larger than the natural Doppler but still small enough such that it can be tracked by the channel estimator at the User Equipment (UE).
Channel estimation is done using the downlink cell specific reference signals (CRS) that are present regardless the UE being attached to the network or not.
FSK was selected due to its robust operation in a fading channel.
We describe the whole AmBC system, use two receivers. Finally, numerical simulations and measurements are provided to validate the proposed FSK AmBC performance.
\end{abstract}

%
%
%

%


\begin{IEEEkeywords}
Ambient Backscatter Communications, LTE Cell Specific Reference Signals, Channel Estimation
\end{IEEEkeywords}

\section{Introduction}
\label{sec:Intro}

\begin{figure*}[t!]
\centering
\includegraphics[trim={9cm 4cm 4cm 5cm},clip=true,width=1.2\columnwidth]{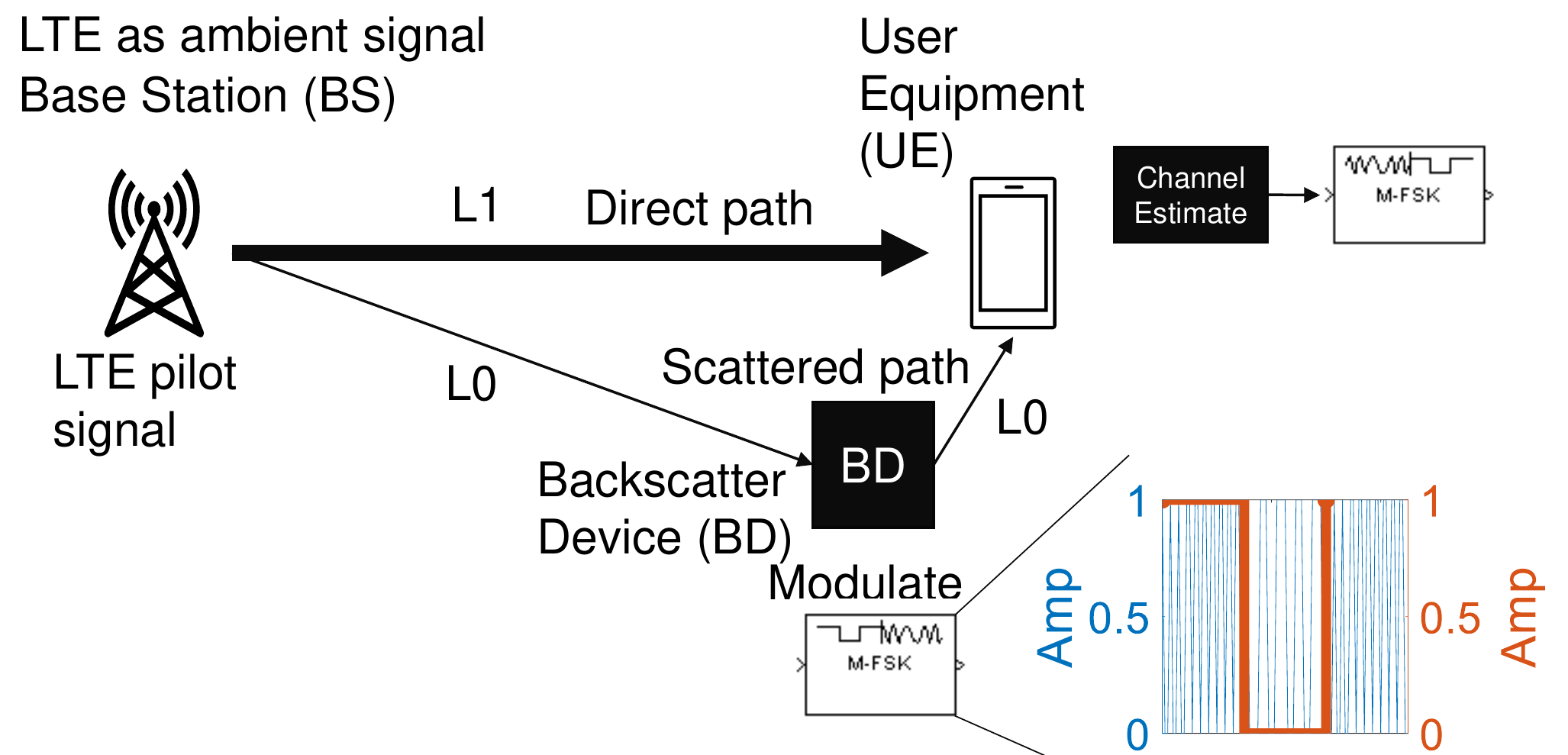}
\caption{ \scriptsize High level structure of the proposed system.
  \label{fig:high_level}}
  \vskip -12pt
\end{figure*}

The introduction of ambient backscatter communications (AmBC) \cite{liu2013ambient} in mobile networks \cite{PhDFara} has recently been proposed for the sustainable development of asset tracking services \cite{9790834}, and to overcome the limitation of radio frequency identification (RFID) based solutions.\

In RFID-based asset tracking \cite{RFIDhandbook} an energy-autonomous and passive RFID tag is illuminated by an RFID reader, with a radio frequency (RF) carrier-wave \cite{1697527}. The tag reflects (backscatters) the wave in a modulated way to send its message, and the reader detects the tag's message in the variations of the backscattered signal. 
As the tag harvests RF energy to power itself, the reader-to-tag range is limited by the reader transmit power to several meters. Tags can therefore be tracked only in places where manual readers or portals are deployed. The short communication range would need to be compensated by increasing the number of readers of portals, but a massive deployment of such devices is not sustainable.\

In comparison, AmBC systems \cite{liu2013ambient} involve three communication nodes instead of only two: an ambient source of RF signals, a backscatter device (BD) and an AmBC receiver device. The BD is similar to a tag. The AmBC receiver reads the BD's message, without having to generate any RF carrier wave, as the BD is already illuminated by the ambient source. In practice, a BD can be implemented with an antenna connected to various matching impedances, through an RF switch driven by a micro-controller. The BD switches between impedances to modulate the reflection according to the message to be transmitted. In \cite{9790834}, it is proposed to use a cellular base station (BS) as an ambient source, and to use a user equipment (UE) as AmBC receiver, to develop a service of asset tracking with ubiquitous coverage. It is almost ``out-of-thin-air": i.e. without generating additional waves, without additional energy, and without deploying massively new equipment such as portals. An energy-autonomous BD harvesting solar energy, called crowd-detectable zero-energy-devices (CD-ZED) is put on the asset to be tracked. Each time the BD (or CD-ZED) gets within few meters of a UE (connected to the cellular network and geo-localised), the BD is detected by the UE and this contact event is reported to the network. Thanks to the anonymous participation of the crowd of UEs, the localisation of the BD is tracked over the cellular network coverage area.
Such CD-ZED concept is one example of the more general category of energy-autonomous devices called zero-energy devices (ZED) \cite{EABZED2020}. Such asset tracking service is one example of ambient internet of things (AIoT) applications, currently being discussed in standardisation for cellular networks \cite{3gppAmbientIoT}. Finally, ZED is one of the key technologies identified for the building of a future and sustainable 6G \cite{9625032}.\

 The CD-ZED concept is applicable to all generations of mobile networks.  Ambient backscatters in 5G networks has been studied in \cite{PhDFara} where it was shown that  BD can be detected by a UE as long as the UE is in the BS coverage and the tag is close to the UE. This is confirmed by successful experiments of ambient backscattering communications conducted with ambient signals from a commercial 4th generation (4G), 5th generation (5G) networks in \cite{9790834}, in very few test locations, far from the BS. 
The previous works \cite{8423609, 6395266} used power detector based receivers that have limited performance due to the high variability of the mobile downlink signals. Very recently, to improve 4G AmBC performance,  \cite{perviousPaper} proposed to use knowledge about pilots of the ambient source (i.e. the BS) at the AmBC receiver (i.e. UE) side. Previously similar approach has been utilized in the context of Wi-Fi standard \cite{10.1145/2934872.2934901}. Unfortunately Wi-Fi pilot transmission is sporadic and thus sub-optimal for reading BD signals. In comparison, LTE pilot signals called cell specific reference signals (CRS) ~\cite{perviousPaper} are always broadcasted by LTE base station.
The CRS structure is standardised and known and used by the UE to estimate the downlink channel. In this paper, we propose to use the UE channel estimator as a receiver for the BD messages.
\par
Performance of AmBC receivers using LTE CRS knowledge is affected by BD modulation method.
In~\cite{perviousPaper}, on-off keying (OOK) modulation was used. That is, BD switched between two load impedances.
Unfortunately, a simple OOK signal occupies frequencies where Doppler components from all the channel paths are also present making it difficult to separate scattered path from the direct path causing so called direct path interference \cite{Biswas21}. In addition, the symbol duration (i.e. switching period) used by BD tend to be long compared to the channel coherence time making the BD signal vulnerable to fast fading.

\par
\par
\par
\textbf{Contributions.}~  
In this paper, we propose to use FSK type modulation in AmBC system. 
For FSK backscatter signal, backscattered path is separated from the direct path in frequency domain, so that the direct path interference  can be cancelled.
BD introduced artificial frequency shift, which we referred to as \emph{frequency key}, is selected to be higher than Doppler effect, and to be lower than channel estimation tracking speed.
Also, the fact that CRS signals are presented only at certain orthogonal frequency division multiplex (OFDM) symbols, limited the frequency key selection. 
Moreover, FSK allows for noncoherent reception that does not depend on the channel parameters. 
The contributions are listed as follows.
\begin{itemize}
  \item 
  The BD signal is generated by the same OOK modulator as in \cite{perviousPaper}, but the generated waveform is selected to approximate FSK. We also discuss square wave FSK that uses rectangular pulses instead of sinusoidal signals.
  We investigate the impact of non-uniform CRS sampling frequency. The FSK frequency keys are carefully selected to cooperate with non-uniform LTE CRS.
  \item 
  The AmBC receiver directly utilizes the channel estimates obtained from LTE CRS pilot signal, instead of full channel state information.
  Two types of receivers, coherent and noncoherent methods are proposed.
  The simulation proves that coherent method outperforms energy detector.
  
  \item Finally, the proposed system is validated by a proof-of-concept implementation and corresponding measurements.
\end{itemize}



The paper is structured as follows: Section \ref{sec:Intro} introduces AmBC, motivation and contribution of this work.
Section \ref{sec:System} describes components of the proposed system.
Section \ref{sec:BC_signal} derives the signal model of the backscatter signal.
Section \ref{sec:LTE_ch} outlines the channel estimation algorithm at the AmBC receiver.
Section \ref{sec:Receiver} presents proposed receiver structure.
Section \ref{sec:Performance} simulates this AmBC system performance and designs a measurement to validate it. 
Finally, a conclusion is draw in Section \ref{sec:Conclusions}.

\section{System description}
\label{sec:System}


\begin{figure}[t!]
\centering
\includegraphics[width=0.85\columnwidth]{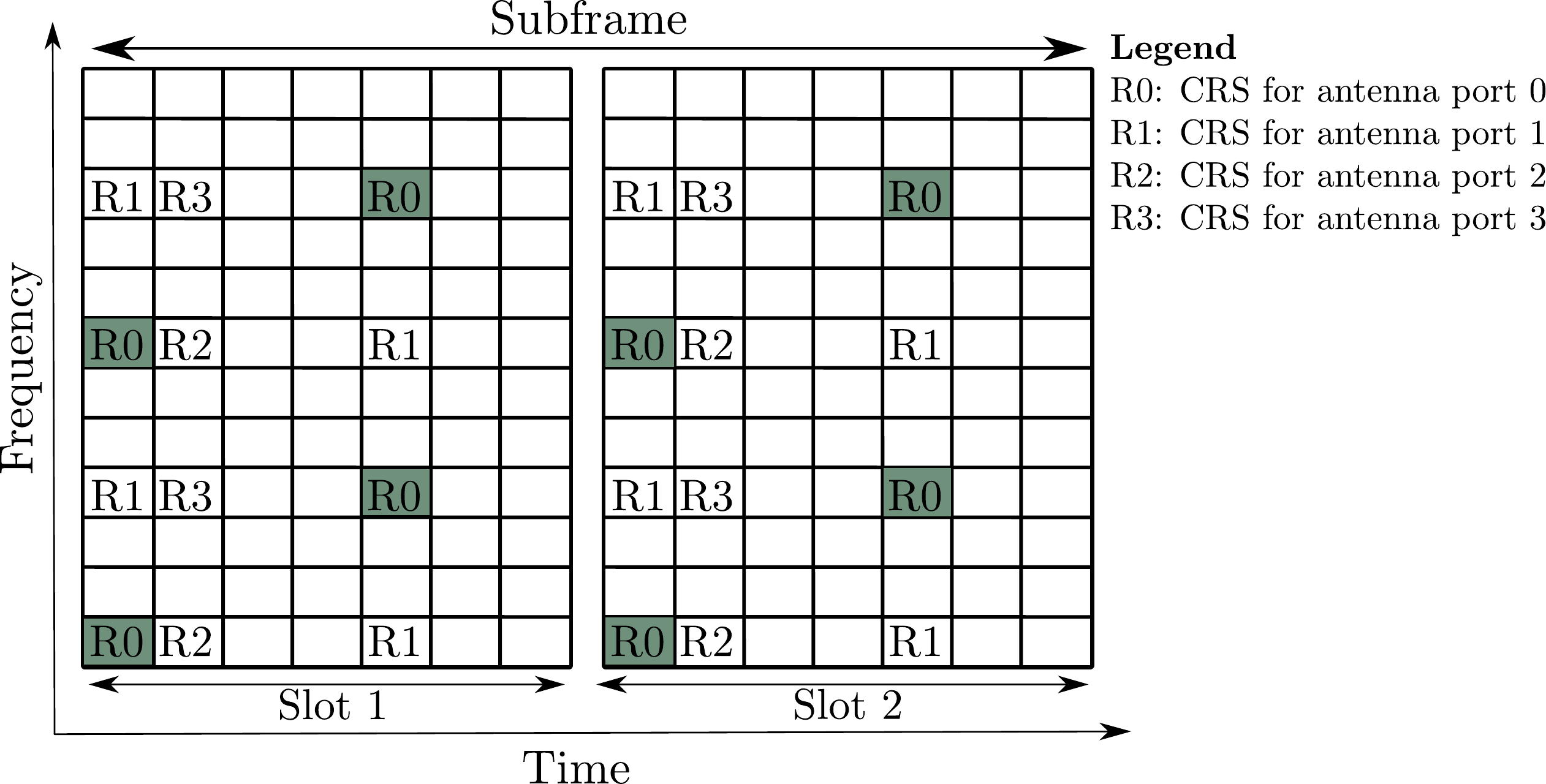}
\caption{ \scriptsize LTE Release 8 Cell Specific Reference Signal for antenna ports 0,1,2, and 3. 
  \label{fig:LTE_CRS}}
  \vskip -12pt
\end{figure}

 We consider an AmBC system consisted of LTE BS (also referred to as NodeB) acting as an ambient source, a UE acting as an AmBC receiver, and a BD as illustrated in Fig. \ref{fig:high_level}.  UE uses the primary and secondary synchronization signals transmitted by NodeB to achieve radio frame, subframe, and symbol synchronicity with the BS. It also identifies the center of the channel bandwidth and deduces the Physical Channel Identity (PCI). After this initial acquisition process, UE uses downlink CRS to estimate the downlink channel. Fig. \ref{fig:LTE_CRS} illustrates CRS for antenna ports 0 to 3. As can be seen from the Fig. \ref{fig:LTE_CRS}, two channel estimations are obtained per 0.5 ms slot leading to 4 kHz non-uniform channel sampling rate. 

BD does not have a pulse shaping filter so it is limited to square pulses which will cause it to have very wide bandwidth.
The square wave frequency shift keying is illustrated as subplot in Fig. \ref{fig:high_level}.
Since our receiver has narrowband, aliasing is unavoidable which causes further challenge for the receiver operation. Furthermore, since we did not implement clock drift compensation, we needed to use noncoherent FSK receiver for the backscatter symbols. A synchronization header is prefixed at the beginning of data packet.

The impulse response of the channel from BS to the AmBC receiver is
\begin{equation}\label{eq:Ch_h}
\begin{aligned}
        h[\tau;t] &= x(t)\sum_{k\in\mathcal{K}_0} a_k[t] \delta[\tau-\tau_k(t)]\\
        &+ \sum_{k\in\mathcal{K}_1} a_k[t] \delta[\tau-\tau_k(t)] ,
\end{aligned}
\end{equation}
where $a_k(t)$, $\tau_k(t)$ are the time varying amplitude and delay of the $k^\textrm{th}$ multipath component and $\delta(\tau)$ denotes the Dirac's delta function.
The bandwidth of the channel tap gain $a_k(t)$ is defined by the Doppler $f_D$ frequency shift in the channel.
In fig \ref{fig:high_level}, the direct path components $\mathcal{K}_1$ from an LTE NodeB to UE is indicated by the thick arrow.
The thin arrow from BS to UE via BD represents $\mathcal{K}_0$ BD modulated scattered components.
\par

\section{Back scattered signal}
\label{sec:BC_signal}

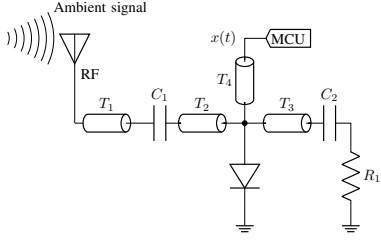
\begin{figure}[t!]
\centering
\resizebox{0.6\columnwidth}{!}{
    \begin{circuitikz}[american]
    \draw (-0.5,0) node[antenna] {RF}
        to [transmission line=$T_1$] (1,0)
        to [capacitor=$C_1$] (2,0)
        to [transmission line=$T_2$] (3,0) -- (4,0)
        to [transmission line=$T_3$] (5,0)
        to [capacitor=$C_2$] (6,0)
        -- (6,-0.5)
        to [resistor, R=$R_1$] (6,-2)
        node [ground]{};
    \draw (3.5,0) to [short, *-] (3.5,-0.5)
        to [diode=$D_1$] (3.5,-2)
        node [ground]{};
    \draw (3.5,0) -- (3.5,0.5)
        to [transmission line=$T_4$] (3.5,1.5);
    \draw (-1,2) node[waves, left]{Ambient signal};
    \draw (3.5,1.5) -- (3.5,2)
        -- (4,2)
        node [msport, circuitikz/monopoles/msport/width=0.75](lable){MCU};
    \node[]()[left of = lable]{$x(t)$};
    \end{circuitikz}
}
\caption{ \scriptsize Circuit diagram of the BD. 
  \label{fig:circuit}}
  \vskip -12pt
\end{figure}

The BD performs load modulation on the incident signal illuminating its antenna, in Fig. \ref{fig:circuit}. That is, it varies its complex antenna load impedance between two states $Z_0$ and $Z_1$. The BD reflection coefficient is given by
$$
\Gamma_x=\frac{Z_x-Z_a^*}{Z_x+Z_a^*}
$$
where $Z_x$ denotes the load impedance in state $x\in\{0,1\}$ and $Z_a$ is the antenna impedance.
In on-off keying case, we would in ideal case have $Z_0=Z_a^*$ and $Z_1=0$ resulting in $\Gamma_0=0$ and $\Gamma_1=-1$.
In practical implementation, the load impedance is switched by a diode as control circuit \cite{8885700}.
In Fig. \ref{fig:circuit}, a micro controller unit (MCU) controls that diode switch, which introduces artificial Doppler.

In the previous work \cite{perviousPaper}, OOK method was introduced for backscatter communication.
The impulse response of channel $h[\tau;t]$ are different when BD in on ($x(t)=1$) or off ($x(t)=0$) status in Eq. \ref{eq:Ch_h}.
\begin{equation*}
\begin{aligned}
        h_\text{on}[\tau;t] &= \sum_{k\in\mathcal{K}_0} a_k[t] \delta[\tau-\tau_k(t)]+\sum_{k\in\mathcal{K}_1} a_k[t] \delta[\tau-\tau_k(t)]\\
        h_\text{off}[\tau;t] &= \sum_{k\in\mathcal{K}_0} a_k[t] \delta[\tau-\tau_k(t)]
\end{aligned}
\end{equation*}
For OOK, ambient signal and Doppler effect strongly influence the channel estimation.
Compared with OOK, FSK shifts the frequency in spectrum and avoids the influence of ambient LTE signal.
FSK shifts RF ambient signal on different frequency key.
It is a good feature so that the frequency keys could be specially designed to compromise Doppler effect and avoid influence of ambient signal. 
\par

The BD aims at causing artificial Doppler that is higher than the natural Doppler in the channel such that the receiver would be able to distinguish between the direct path components (multipath components in $\mathcal{K}_1$) and BD modulated scattered components (multipath component in $\mathcal{K}_0$).
BD does this by generating periodic rectangular wave 
$\tilde{x}_k(t)=\tilde{x}_k(t+T_k)$:
$$
\tilde{x}_k(t) =\sum_{n=-\infty}^\infty \mathsf{rect}\left[\frac{2(t-nT_k)}{T_k}\right], \quad k=0,1
$$
where $\mathsf{rect}(t)$ is the unit rectangular pulse and the index $k$ indicates whether bit 0 or 1 was transmitted.
$\tilde{x}_0(t)$ and $\tilde{x}_1(t)$ are sparse or tight square waves with infinite extension.
However, the BD symbol duration is $T_{BC}$, which is the red line in the subplot of Fig. \ref{fig:high_level}.
Hence the generated BD pulse is
$$
x_k(t) = \mathsf{rect}\left(\frac{t}{T_{BC}}\right)\tilde{x}_k(t), \quad k=0,1
$$
In time domain, $x_0(t)$ or $x_1(t)$ look like blue line segments in the subplot of Fig. \ref{fig:high_level}.
The Fourier transform of the BD symbol is given by
$$
X_k(f)=\frac{T_{BC}}{2}\sum_{l=-\infty}^{\infty} \sinc\left(\frac{1}{2}l\right)\sinc\left[\left(f-\frac{l}{T_k}\right)T_{BC}\right]
$$
where $\sinc(x)=\sin(\pi x)/(\pi x)$.
The harmonics of the rectangular wave nominal frequency $l\frac{1}{T_k}$, $l=3,5,...$ attenuate slowly implying that the square wave has a wide bandwidth.

\section{LTE channel estimation}
\label{sec:LTE_ch}

In the LTE system, the BS transmits CRS in every subframe. Fig.~\ref{fig:LTE_CRS} illustrates the CRS allocation for antenna ports 0 to 3 when the system is using normal cyclic prefix. LTE networks may operate using different CRS configurations. In a non-shifted CRS configuration, all cells use the same CRS time and frequency resources. In a shifted CRS configuration, different cells transmit CRSs on resources that are shifted in frequency. The non-shifted configuration avoids CRSs interference on data transmissions, but is also associated with a systematic CSI estimation error; especially noticeable at low traffic. Using the shifted configuration, the CRSs interfere with data transmissions, but the CSI estimation error is smaller \cite{madhugiri2014}. In this paper, we consider the case of shifted CRS, and the AmBC receiver uses pilots transmitted from antenna port 0.

In case of normal cyclic prefix, the OFDM symbol duration in LTE is $T_s =71.4 \; \mathrm{\mu s}$ except for the symbol 0 that has longer prefix.  Also since pilots are only present in symbols 0 and 4, we get irregular sampling of the channel.
Let $T_{slot}=0.5\;\mathrm{ms}$ denote the slot length
and let $ \Delta T = 4T_s - \frac{T_{slot}}{2} =35.6 \;\mathrm{\mu s} $ denote the offset of the second channel sampling instant compared to regular sampling interval $T_r=T_{slot}/2$.
If samples in regular sampling interval $T_r=0.25\;\mathrm{ms}$, that would lead to 4 kHz sampling frequency. 

From the transmitted CRS, we obtain frequency domain channel estimates $\hat{H}[n;t]$ for time instants $t\in\{0,T_{slot}/2+\Delta T, T_{slot}, 3/2T_{slot}+\Delta T,\cdots\}$ during which pilots were send. $n$ is discrete frequency. Assuming that the channel stays approximately constant during the transmission of single OFDM symbol, inverse Fast Fourier Transform of $H[n;t]$ at time instants $t$ yields the following channel taps
\begin{eqnarray*} \label{eq:hath}
\hat{h}[l;t] &=&x(t)\sum_{k\in\mathcal{K}_0} a_k^b(t)\mathsf{sinc}\left(l-\tau_k(t)W\right)\\
&+&\sum_{k\in\mathcal{K}_1} a_k^b(t)\mathsf{sinc}\left(l-\tau_k(t)W\right)
+z_l(t),
\end{eqnarray*}
where $W=\frac{1}{T_s}$ denotes the utilized bandwidth, $a_k^b(t)=e^{-i2\pi f_c \tau_k(t)}a_k(t)$ denotes the baseband equivalent channel tap of the $k^\textrm{th}$ multipath component, $f_c$ is the carrier frequency, and $z_l(t)$ denotes the the estimation noise, AmBC signal $x(t)$ could be $x_0(t)$ or $x_1(t)$. The LTE system is synchronized to the shortest path which appears in the first channel tap. Backscattered signal component is likely to be much smaller than the direct path component leading to very small signal-to-noise ratio (SNR). As a consequence, the distance between BD and receiver is short in most practical deployments and thus most of the BD scattered power would be in the first channel tap. 
Hence, in the receiver it is sufficient to find just
\begin{equation*}
    \hat{h}[0;t] =x(t)h_0(t)+h_1(t),
\end{equation*}
where $h_0(t)$ and $h_1(t)$ contain the scattered and direct path components that appear in the first channel tap after sampling.

Let ${s}(t)=\delta(t)+\delta(t-T_{slot}/2-\Delta T) $ 
be periodic sampling signal ${s}(t+T_{slot})={s}(t)$ where $T_{slot}$ denotes 
the slot length and $\Delta T$ denotes the time offset of the second pilot in the slot compared to half of the slot time $T_{slot}/2$.
Since $s(t)$ is periodic, we can express it in terms of Fourier-series as
\begin{eqnarray*}
s(t)=\sum_{l=\infty}^\infty s_l e^{i2\pi\frac{t}{T_{slot}}l}
\end{eqnarray*}
where the Fourier series coefficient are given by
\begin{eqnarray*}
s_l &=& \frac{1}{T_{slot}} \int_0^{T_{slot}} s(t) e^{-i2\pi\frac{t}{T_{slot}}l}dt\\
&=& \frac{1}{T_{slot}} \left(1+e^{-i\pi\left(1+2\frac{\Delta T}{T_{slot}}\right)l} \right)\\
&=&\frac{1}{T_{slot}} \left(1+(-1)^l e^{-i2\pi \frac{\Delta T}{T_{slot}}l} \right)\\
&=& \frac{2}{T_{slot}} \frac{1+(-1)^l}{2} +\frac{1}{T_{slot}} (-1)^l \left(e^{-i2\pi \frac{\Delta T}{T_{slot}}l } -1. \right)
\end{eqnarray*}

The sampled channel is $h_s(t)=\hat{h}[0;t]s(t)$.  Now using the Fourier series representation of $s(t)$ and taking the Fourier-transform of $h_s(t)$, we obtain the
Discrete Time Fourier Transform (DTFT) of the sampled channel response:
\begin{eqnarray*}
H_s(f)&=&\frac{2}{T_{slot}}\sum_{l=-\infty}^{\infty}H\left(f-\frac{2l}{T_{slot}}\right) \\
&+& \frac{1}{T_{slot}}\sum_{l=-\infty}^{\infty}\varepsilon_l H\left(f-\frac{l}{T_{slot}}\right)
\end{eqnarray*}
where $\varepsilon_l=(-1)^l  \left(e^{-i2\pi \frac{\Delta T}{T_{slot}}l } -1 \right)$.

The first (upper) sum corresponds to the spectrum of the channel sampled at rate $\frac{2}{T_{slot}}= 4$ kHz and the second (lower) sum contains additional aliased components due to the irregularity of the sampling $\Delta T$. The figure \ref{fig:DTFT} shows that after sampling, the spectrum contains the desired FSK signal, its harmonic components as well as aliased harmonics.

Even with 4 kHz sampling frequency, we would experience severe aliasing of the harmonic components of the square waves. Due to irregular sampling, we will see additional aliased components, but they are attenuated by the factor $|\varepsilon_l|$. To be on the safe side, we select the square wave nominal frequencies be on the range $f_k\in[200, 1000]$ Hz. The lower limit is selected to be larger than the natural Doppler in the channel such that the direct path $h_1(t)$ can be filtered away using high-pass filter. The upper frequency is selected to be small enough to avoid additional aliasing due to irregular sampling.  

Even if the two backscatter symbols $x_0(t)$ and $x_1(t)$ would have been selected to be orthogonal, after sampling they will interfere with each other. Due to aliasing, it turns out that orthogonal choices $f_1=Kf_0$ for integer $K$ leads to high interference from aliased harmonics hitting the other symbol. It is thus seems advantageous to take $K$ not to be integer. 

\section{Receiver structure}
\label{sec:Receiver}
\begin{figure}[t!]
  \includegraphics[width=\columnwidth]{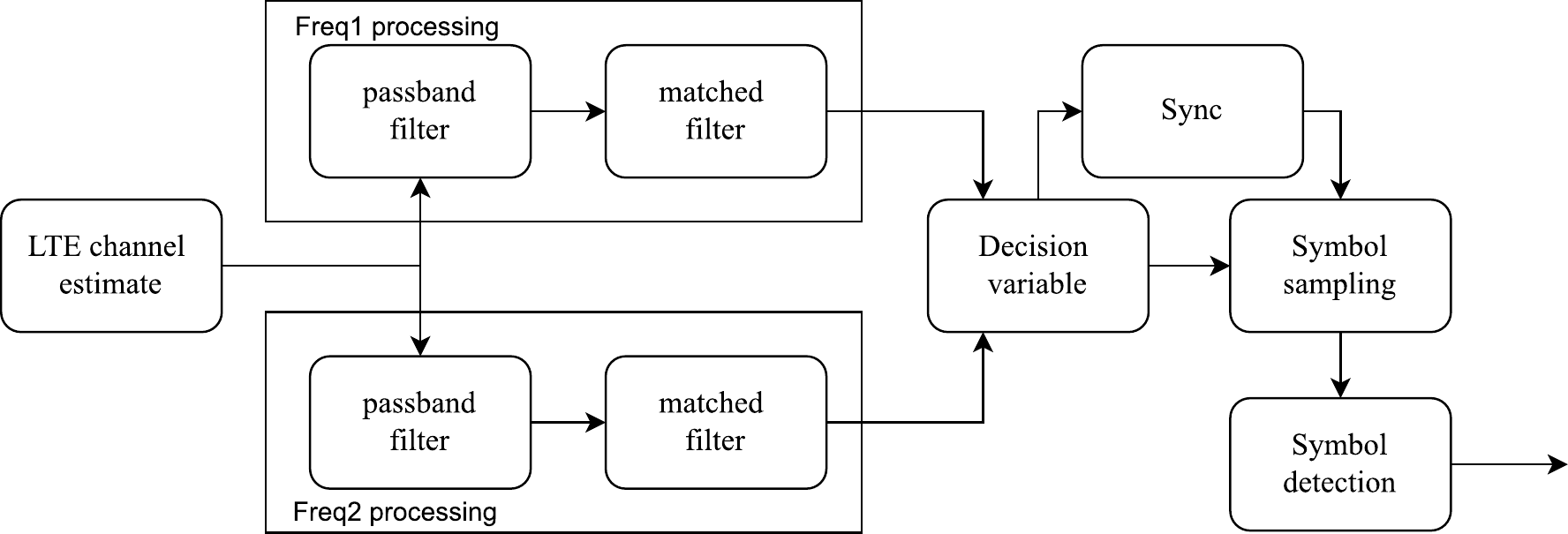}
  \caption{Flow chart of the proposed backscatter receiver.}
  \label{fig:RecFlowchart}
   \vskip -12pt
\end{figure}

The flow chart in Fig. \ref{fig:RecFlowchart} shows the algorithm steps of the proposed backscatter receiver.
In this section, the purposes of some steps in receiver are elaborated on.
\subsection{Band-pass Filter}
It easy to assume BD symbol keeps frame synchronicity when received and demodulated by UE.
$m\in \mathbb{N}$ is defined as the $m$-th symbol backscatter sending at time $t=mT_{BC}$.

Channel phase of scatter path $\mathrm{arg}\{g_0(mT_{BC})\}$ is ambiguous due to synchronization.
Power of the first channel tap $l=0$ does not contain the phase uncertainty.
The receiver only operates with channel tap power $y[m]=|\hat{h}[0;m]|^2$.

If we don't consider noise in the channel, the channel power approximately satisfied the following relationship
\begin{equation*}
    y[m]\approx x[m]\beta[m] + \alpha[m],
\end{equation*}
where $x[m]$ is the BD signal,
$\alpha[m]=|h_1(mT_{BC})|^2$, and $\beta[m]=|h_0(mT_{BC})|^2+2{\mathrm{Re}}\{h_1^*(mT_{BC})h_0(mT_{BC})\}$,
considering the fact that $x^2[m] = x[m]$.
\par
To separate this two channels, a high-pass filter and a low-pass filter is required.
High pass is designed to block $\alpha[m]$. And low-pass aims to constrain the harmonic frequency and other interference.
In practice, they can be combined as a band-pass filter (BPF).
Considering the frequency keys of FSK are only several hundreds Hz, Doppler effect is the principle thereat of propose backscatter receiver.
Doppler effect and frequency drift of BS and UE contribute to the channel change of $\alpha[m]$ and $\beta[m]$ in a small time scale.
By switching the BD at a higher frequency than the maximum Doppler in the channel, the Doppler frequency is restrained by filter.
With help of a high-pass filter on $y[m]$ to remove the direct path interference $\alpha[m]$, BD modulated path $\beta[m]$ component is distinguished in frequency domain.
Thus, two frequency keys of the BD FSK symbols left.
In the base band, FSK symbol is designed two frequency keys, namely $f_0$ and $f_1$.
$$f_k=1/T_k,\quad k=0,1$$
\par
\subsection{FSK Demodulator}
\label{subsec:FSK Demodulator}
After passing a high-pass filter and low-pass filter, received 2-FSK signal $y_f[m]$ is demodulated. We propose both coherent and noncoherent, power detection based method for the task in Fig. \ref{fig:DemodSharedStep}.
\par

\begin{figure}[t!]
\centering
\includegraphics[width=0.8\columnwidth]{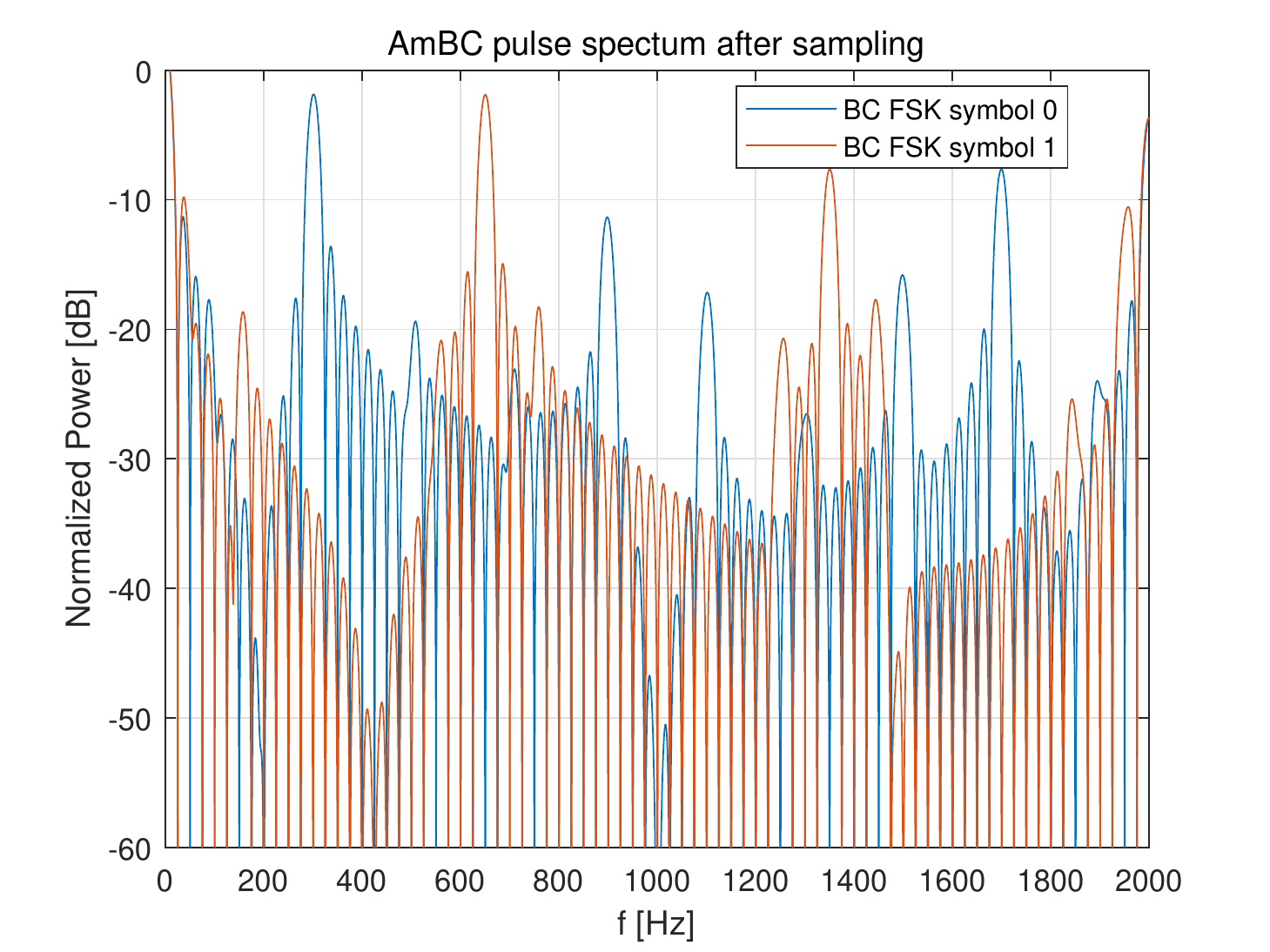}
\caption{ \scriptsize Discrete Time Fourier Transfrom of the sampled FSK signal. 
  \label{fig:DTFT}}
  \vskip -12pt
\end{figure}
Both of them share one step, that is, filtering $y_f[m]$ at $f_0$ and $f_1$ firstly.
There are aliasing effect of two FSK as Fig. \ref{fig:DTFT} illustrated.
Harmonic components of one FSK key could unfortunately hit another FSK key.
A BPF is applied to exclude others frequency leakage and to constrain interference frequency.
Denote $y_f[m]$ pass BPF at center frequency $f_0$, as $y_\text{f0}[m]$, and $y_f[m]$ pass BPF at center frequency $f_1$, as $y_\text{f1}[m]$.
\par
\begin{figure}[t!]
\centering
\includegraphics[width=\columnwidth,trim=2cm 10.5cm 11cm 11cm,clip]{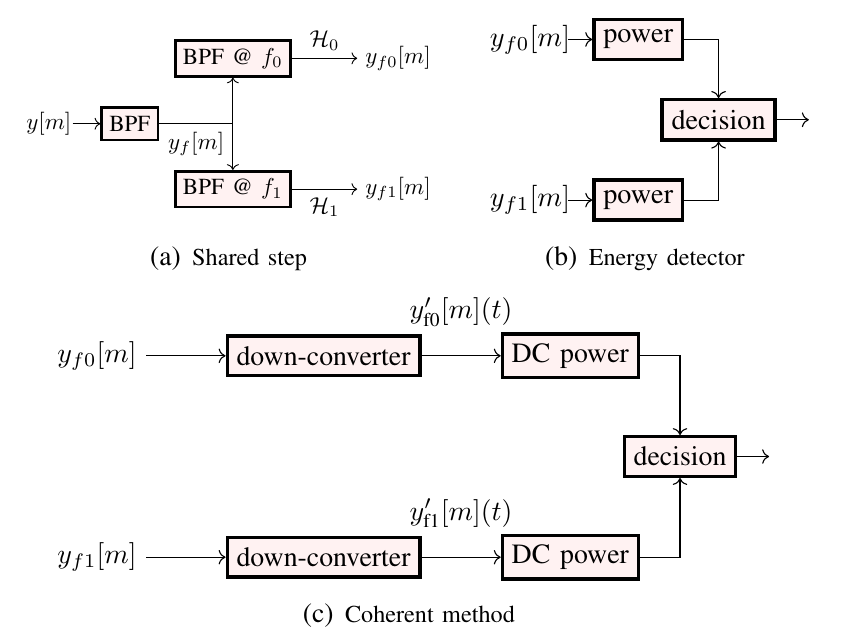}
\caption{ \scriptsize Flowchart of the FSK demodulator. We proposed both coherent and noncoherent detectors.
Flow chart (a) is shared step of both detectors. Flow char (b) and (c) are energy detector and coherent method separately.
Complete energy detector is composed by (a) and (b). Also, total coherent method consists of (a) and (c).
  \label{fig:DemodSharedStep}}
  \vskip -12pt
\end{figure}
Energy detector compares the power of specturm $f_0 \pm \Delta f$ and power of spectrum $f_1 \pm \Delta f$.
An FSK symbol is decided based on the frequency spectrum which contains higher power.
For FSK symbol $x[m]$, hypothesis testing $\mathcal{H}_0$ denotes that backsactter device sends $x[m]$ symbol 0.
Similarly, hypothesis testing $\mathcal{H}_1$ refers that $x[m]$ is symbol 1.
So for energy detector,
\begin{equation*}
   E\left[ \left|y_\text{f1}[m](t)\right|^2 \right]
   \mathrel{\substack{\mathcal{H}_0\\ \lessgtr \\ \mathcal{H}_1}}
   E\left[\left|y_\text{f0}[m](t)\right|^2\right]
\end{equation*}
\par

Coherent method down convert the signal to base band by multiplying the signal with $\exp\left(-j2\pi f_c t\right)$.
\begin{equation*}
\begin{aligned}
    y'_\text{f0}[m](t) = y_\text{f0}[m](t)e^{-j2\pi f_0 t}\\
    y'_\text{f1}[m](t) = y_\text{f1}[m](t)e^{-j2\pi f_1 t}
\end{aligned}
\end{equation*}
The decision is made by comparing the base band power of $y'_\text{f0}[m](t)$ and $y'_\text{f1}[m](t)$.
The base band power is in fact the power of average signal.
And then, the FSK symbol decision is make based on the power comparison:
\begin{equation*}
   \left|E\left[y'_\text{f1}[m](t)\right]\right|^2
   \mathrel{\substack{\mathcal{H}_0\\ \lessgtr \\ \mathcal{H}_1}}
   \left|E\left[y'_\text{f0}[m](t)\right]\right|^2
\end{equation*}

\section{Simulation and Validation}
\label{sec:Performance}

\subsection{Parameters}
\label{subsec:Parameters}
We propose to select FSK $f_0=300$ Hz and $f_1=650$ Hz, and $T_{BC}=40$ ms which corresponds to six periods of the symbol `0' wave, and 13 periods of the symbol `1' wave.
A 200 Hz bandwidth BPF is applied in FSK demodulator with different center frequency.
$y_\text{f0}[m]$ is filtered by a bandwidth BPF 200 Hz at 300 Hz center frequency and $y_\text{f1}[m]$ is filtered by a BPF bandwidth 200 Hz at 650 Hz center frequency.
\par
The ambient signal is LTE CRS signal, whose paramters are given below. Frequency Division Duplex (FDD) operation is assumed. BS, uses only antenna port 0. That is, it has only single antenna. Also the UE is assumed to be equipped only with a single antenna.
The utilized downlink carrier frequency is 486 MHz with bandwidth 7.68 MHz which accommodates 21 resource blocks. 
\par

\subsection{Simulations}
\label{subsec:Perf_sim}

\par
Free-space path loss (FSPL) model are applied on different propagation paths with additive white Gaussian noise (AWGN) added in the channel. For backscattered channel, the pathloss is the product of the two links: from transmitter to BD and from BD to receiver \cite{griffin2009complete}.
No Doppler effect is considered in this simulation.
The channel impulse response $h(t)$ and FSPL is supposed to obey
\begin{equation}
    \label{eq:fspl}
    L=\left(\frac{4\pi D}{\lambda}\right)^2=E\left[|h(t)|^2\right],
\end{equation}
where $D$ is distance between Tx and Rx, $\lambda=c/f_c$ is wave length of carrier frequency.
The path loss $L$ here is in linear scale.
In this simualtion, the distance from Tx to Rx is 125 m, distance from Tx to BD is 130 m and distance from BD to Rx is 10 m.
In the simulation we consider two propagation paths are both non-line of night (NLOS), obey Rayleigh distribution.
\begin{equation}
    y(t)=h_0(t)s(t)+R_\text{on}h_1(t)s(t)x(t) +n(t), \label{eq:multipath}
\end{equation}
where $R_\text{on}$ is the reflection coefficients of BD under `on' status,
$h_0(t)$ and $h_1(t)$ are channel impulse response of scattered path and direct path,
$s(t)$ is LTE CRS ambient signal and $n(t)$ is AWGN.
Depending on symbols AmBC sent, $x(t)$ can be either $x_0(t)$ or $x_1(t)$.
Footnote 0 refers to direct path between Tx and Rx, and footnote 1 refers to backscatter communication path between Tx and Rx via BD.
In ideal backscater, the reflection coefficient is $\Gamma=-1$ when BD in `on' state and $\Gamma=0$ in the 'off' state, but in practice these values tend to be closer to each other. 

To model non-idealities, we assume that the BD attenuates the reflected signal power by $-20\log_{10}(R_\text{on})=6$ dB. 
\par

\begin{figure}[t!]
    \centering
    \includegraphics[width=0.85\columnwidth]{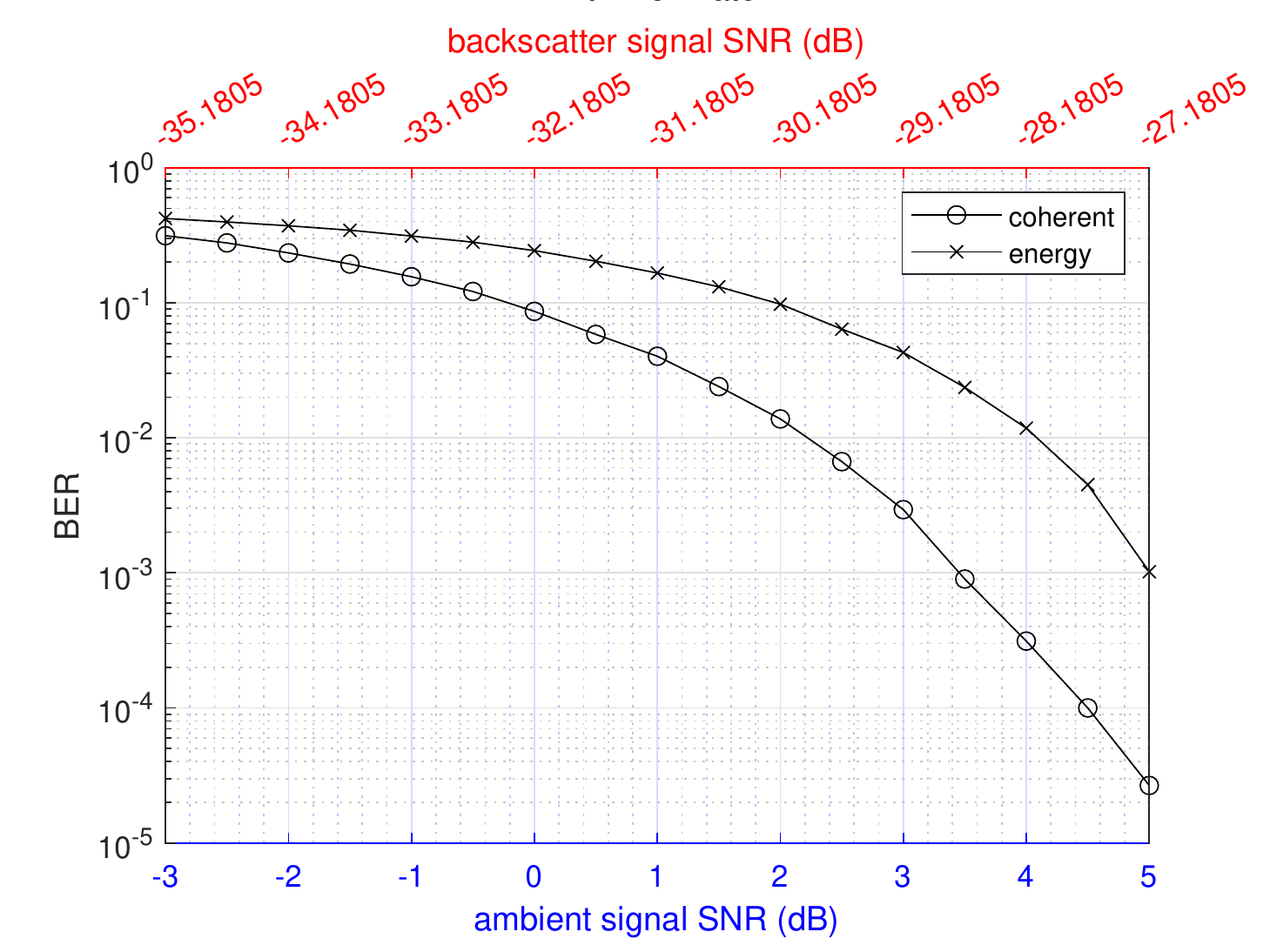} 
    \caption{Theoretical BER of AmBC signal, based on coherent and energy detector.}
    \label{fig:BER_sim}
    \vskip -12pt
\end{figure}

In Fig. \ref{fig:BER_sim}, there are two SNR defined.
Red and blue x-axis are both SNR in dB scale.
The only difference is the definition of signal power in SNR.
$n(t)\sim\mathcal{CN}(0,\sigma_n^2)$.
The power of noise is $P_n=\sigma_n^2.$

The blue x-axis on the bottom is based on CRS power and the red axis at the top is for the received BD modulated signal power.


The power of the CRS is
\begin{equation*}
P_{s1}=E\left[|h_0(t)s(t)+R_\text{on}h_1(t)s(t)x(t)|^2\right]
\end{equation*}
which corresponds to the LTE Reference Signal Received Power (RSRP) in the absence of noise.

Hence for the blue x-axis, we have
\begin{equation*}
    \text{SNR}_1=\frac{P_{s1}}{P_n}
    =\frac{E\left[|h_0(t)s(t)+R_\text{on}h_1(t)s(t)x(t)|^2\right]}{\sigma_n^2}.
\end{equation*}
\par
The red x-axis on the top treats backscatter FSK signal from backscatter as the signal of interest.
The backcatter signal SNR of AmBC is defined as
\begin{equation*}
    \text{SNR}_2
    =\frac{E\left[|R_\text{on}h_1(t)s(t)x(t)|^2\right]}{\sigma_n^2}.
\end{equation*}
As Eq. (\ref{eq:multipath}) illustrating, the power of two path difference is exactly
\begin{equation*}
\begin{aligned}
    10\log{\left(\Delta L\right)} &= 10\log E\left[|h_0(t)|^2\right] - 10\log E\left[|h_1(t)|^2R_\text{on}^2\right]\\
    &=10\log L_0 - 10\log L_1 - 20\log R_\text{on}.
\end{aligned}
\end{equation*}
\par

Using MATLAB LTE toolbox, LTE fading channel model is applied.
Doppler frequency shift is not considered in this simulation, although Doppler effect influences a lot in practice.
MIMO channel propagation is also not setup, because transmitter and receiver antenna is assumed to be SISO.
LTE downlink channel estimator estimates the channel based on that CRS signal.
None OFDM symbol is interpolated between CRS pilots.
\par
The coherent detector algorithm and energy detector algorithm are discussed in subsection \ref{subsec:FSK Demodulator} FSK demodulator.
To smooth the simulation BER curve, we simulate various times of experiments.
High BER points ($\text{BER}>0.01$) simulate 10000 times Monte Carlo experiments.
Low BER points ($\text{BER} \leq 0.01$) simulate 100000 times Monte Carlo experiments.
\par
Simulation uses backscatter communication parameters as subsection \ref{subsec:Parameters} Parameters mentioned.
Fig. \ref{fig:BER_sim} is the results of simulation.
The energy detector is always worse than coherent detector.
Under low SNR, such as -3 dB, the two methods have similar performance.
The BER difference of two FSK demodulators is small.
But the performance at SNR = 5 dB, coherent detector is better one order of magnitude than energy detector.

\subsection{Backscatter signal synchronization}
\label{subsec:Synchronization}
\par
\begin{figure}[t!]
    \centering
    \includegraphics[width=0.95\columnwidth]{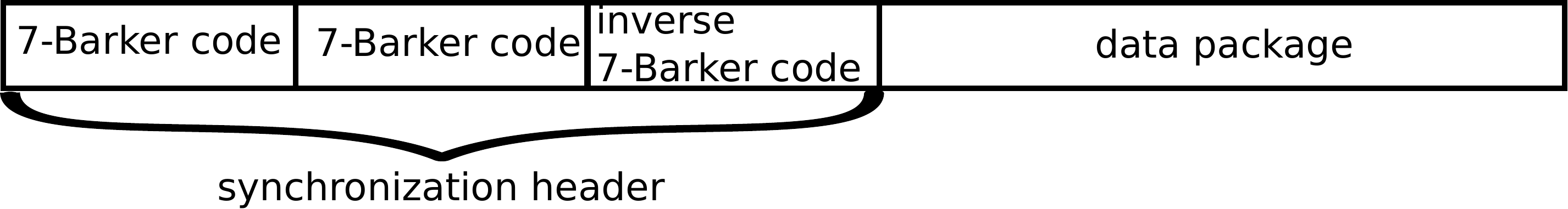}
    \caption{Backscatter frame format of the proposed backscatter.}
    \label{fig:data_format}
    \vskip -12pt
\end{figure}
A special backscatter frame structure is designed to find out the beginning of a backscatter signal.
Backscatter signal is synchronized by three sequences of 7-Barker code.
As Fig. \ref{fig:data_format} showing, there are two parts in one backscatter frame, synchronization header and data packet.
At the beginning of one packet, two continuous sequences of 7-bit Barker code (`0000110') followed by an inverse 7-bit Barker code (`1111001') compose a synchronization header.
Then data packet attaches the synchronize header.
\par
Between two backscatter packets, there is a short period that no FSK symbol is sent, called sleep period.
During sleep period, ambient signal is not shifted and BD kept `off' state.

Compared to previous work \cite{perviousPaper}, clock signal to synchronize is canceled in the proposed method.
Sometimes, backscatter packets would be synchronized incorrectly.
In that case, the data bit error could be incredibly high (over than one third).
The synchronization header bits is already known, as a part of backscatter communication protocol.
By comparing known synchronization bits with demodulated bits, we can evaluate the quality of Rx received data and decide whether synchronization was successful or not. If the measurement indicates that the synchronization failed, we discard the whole packet.
\subsection{Measurements}
\label{subsec:Perf_meas}
This measurement is a validation of aforementioned simulation.
Parameters are same as subsection \ref{subsec:Parameters} Parameters mentioned.
\par
The transmitter is Rhode\&Schwarz (R\&D) SMBV100 signal generator with LTE signal generator packet. 
The generator emits standard LTE signal with 50 resource block (7.68 MHz bandwidth) at 486 MHz carrier frequency and transmission power is 15 dBm.
The frame structure is for SISO system with corresponding synchronization signal and pilots for cell ID 3.
\par
The BD node is an in house design BD as Fig. \ref{fig:circuit}.
Control signal from MCU is driven by RaspberryPi nano, an RP2040-based microcontroller board.
\par
The receiver is a universal software radio peripheral (USRP), connected to a laptop.
Some post signal processing is executed on that laptop, using MATLAB.
\subsubsection{Wired measurement}
\begin{figure}[t!]
    \centering
    \includegraphics[width=0.82\columnwidth]{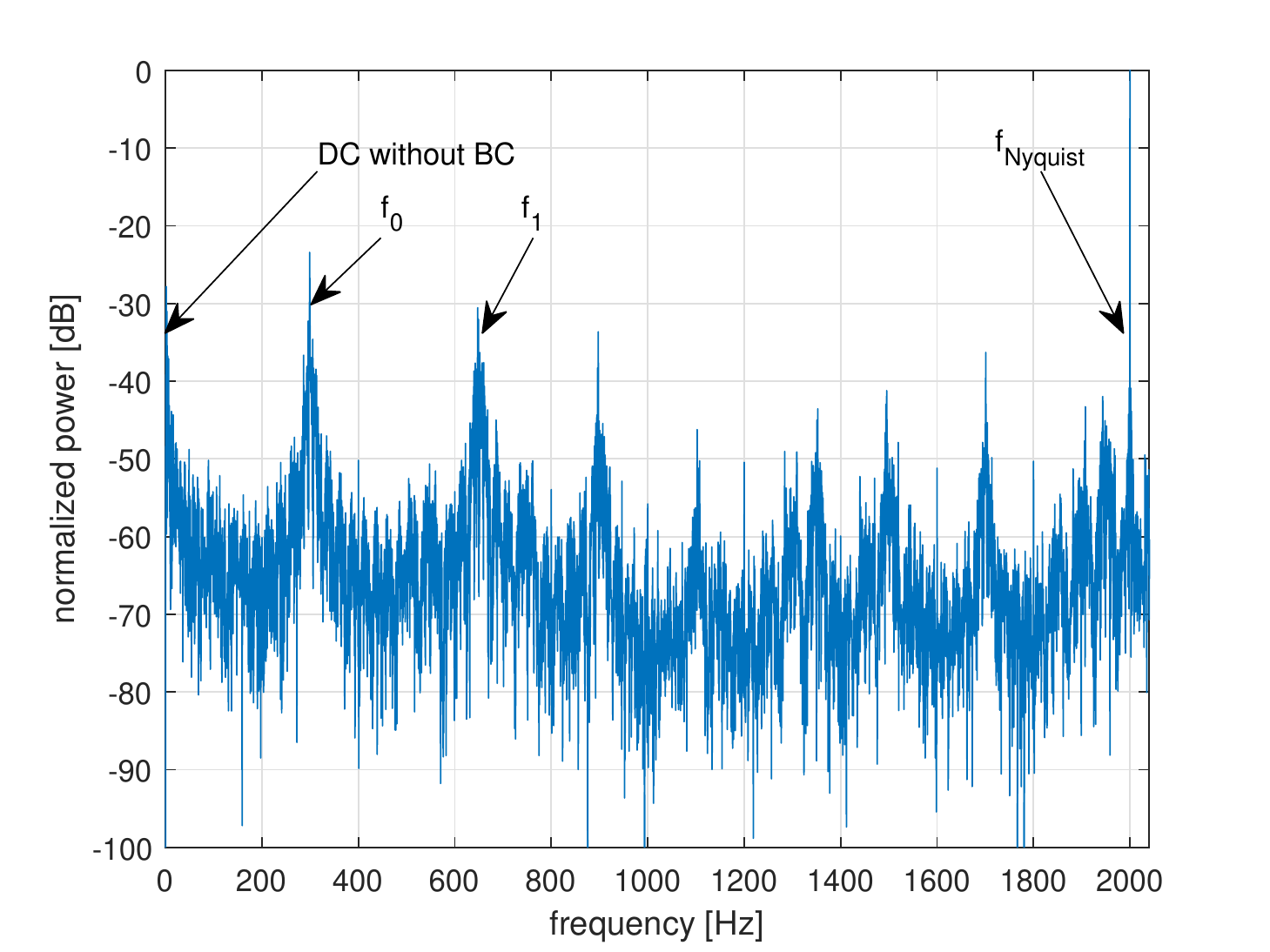}
    \caption{Wired measurement of a FSK modulated BD signal in spectrum.}
    \label{fig:Spectrum}
    \vskip -12pt
\end{figure}
\begin{figure}[t!]
    \centering
    \includegraphics[width=0.85\columnwidth]{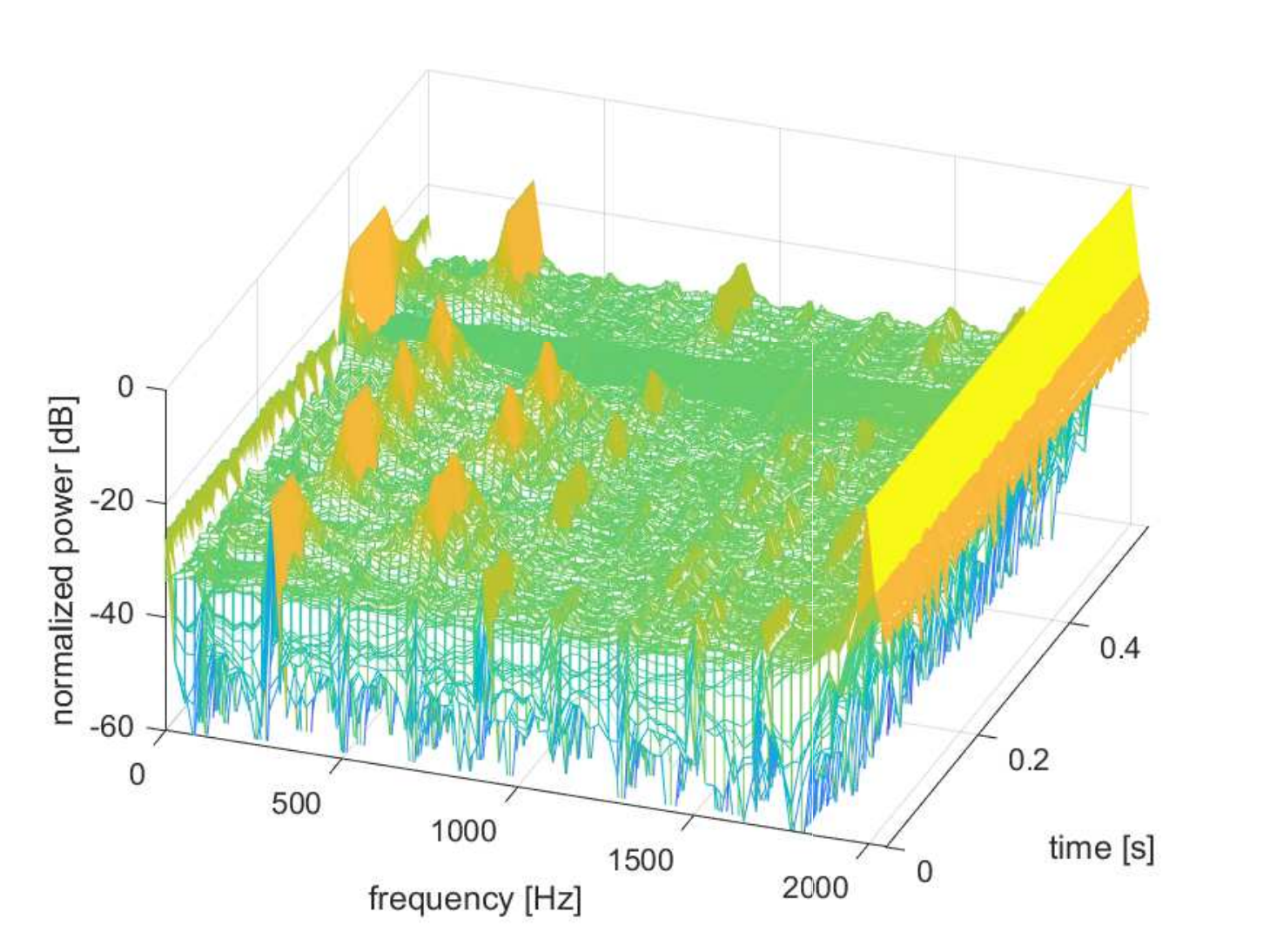}
    \caption{Wired measurement of a FSK modulated BD signal in spectrogram.}
    \label{fig:Spectogram}
    \vskip -12pt
\end{figure}
This experiment measures over the cables in the absence of direct path component $h_1(t)$.
A circulator routes signal from LTE signal generator to BD and then from BD to USRP.
Fig. \ref{fig:Spectrum} and \ref{fig:Spectogram} simply give the spectrum and spectrogram of USRP receiver, respectively.
\par
The two symbols are clearly visible in the spectrum as well as their aliased harmonic components.
In addition there is a strong DC component and a component at 2 kHz corresponding to the uniform sampling frequency $1/T_{slot}$, which arrows $f_\text{Nyquist}$ points to in Fig. \ref{fig:Spectrum}.
The spectrum was obtained using Fast Fourier Transform directly on the measured channel samples $\hat{h}[0,t_s]$ without compensating for the irregularity of the underlying sampling process.
From the spectrogram that illustrates how the spectrum changes in time, we can clearly see the transmitted symbol sequence `11001010' by observing the power at the frequencies $f_0$ and $f_1$.
As illustrated in Fig. \ref{fig:Spectrum} around 0.5 s, there is a 100 ms sleep period, presenting a peak at direct current (DC) component.
\par
The specterogram of square-wave FSK is illustrated as Fig \ref{fig:Spectogram}.
There are two frequency keys $f_0$ (300 Hz) and $f_1$ (650 Hz) appear alternately.
The peak of 2 kHz is caused by uniform sampling frequency $1/T_{slot}$, same as Fig. \ref{fig:Spectrum}.
Other peaks in the spectrum, such as 900 Hz, is caused by alias from other components on the spectrum or harmonic frequency.

\subsubsection{Wireless measurement}
\begin{figure}[t!]
\centering
\includegraphics[width=0.82\columnwidth]{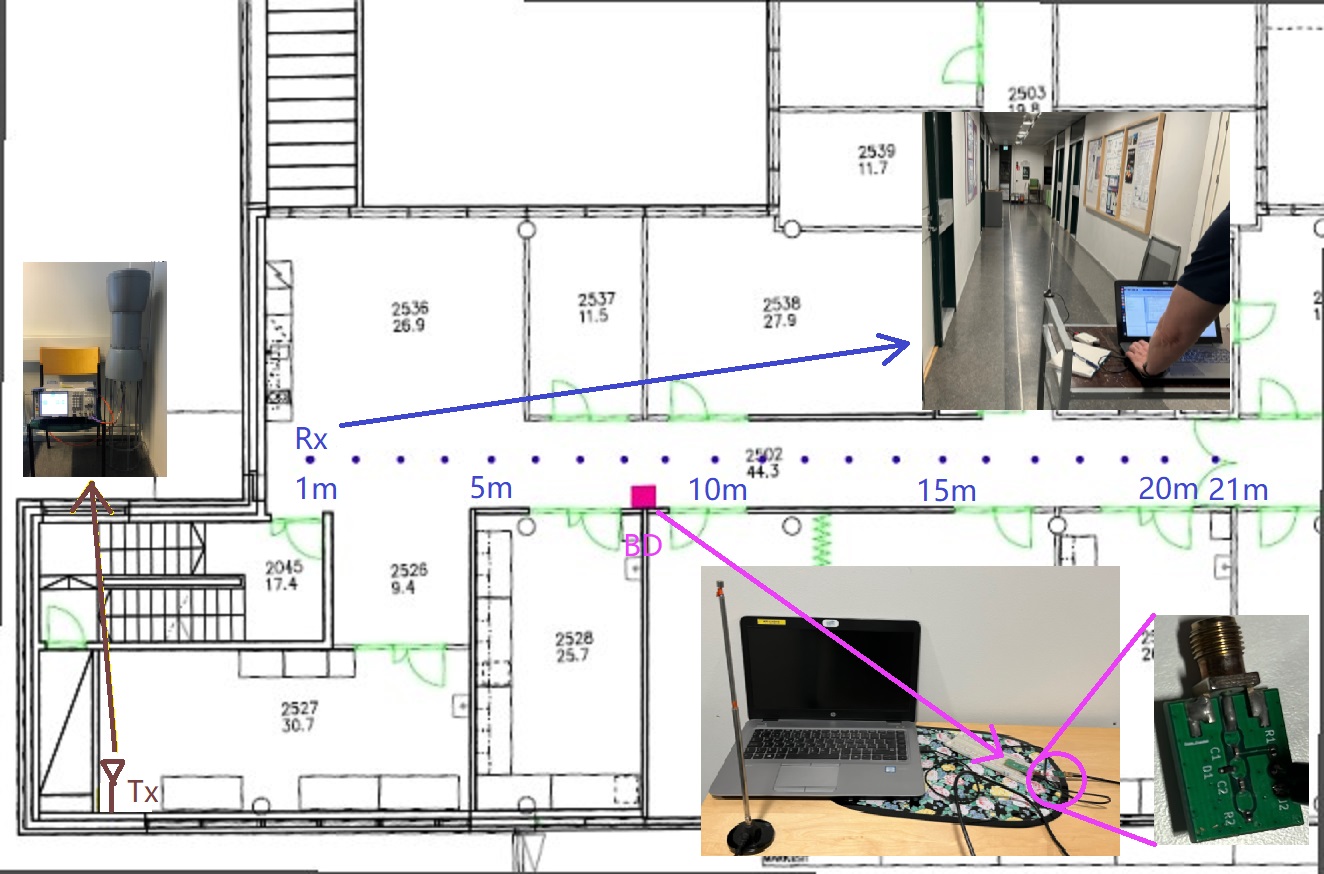} 
\caption{Wireless measurement devices and site floor plan.}
\label{fig:measurement_environment}
\vskip -12pt
\end{figure}

\begin{figure}[t!]
\centering
\includegraphics[width=0.85\columnwidth]{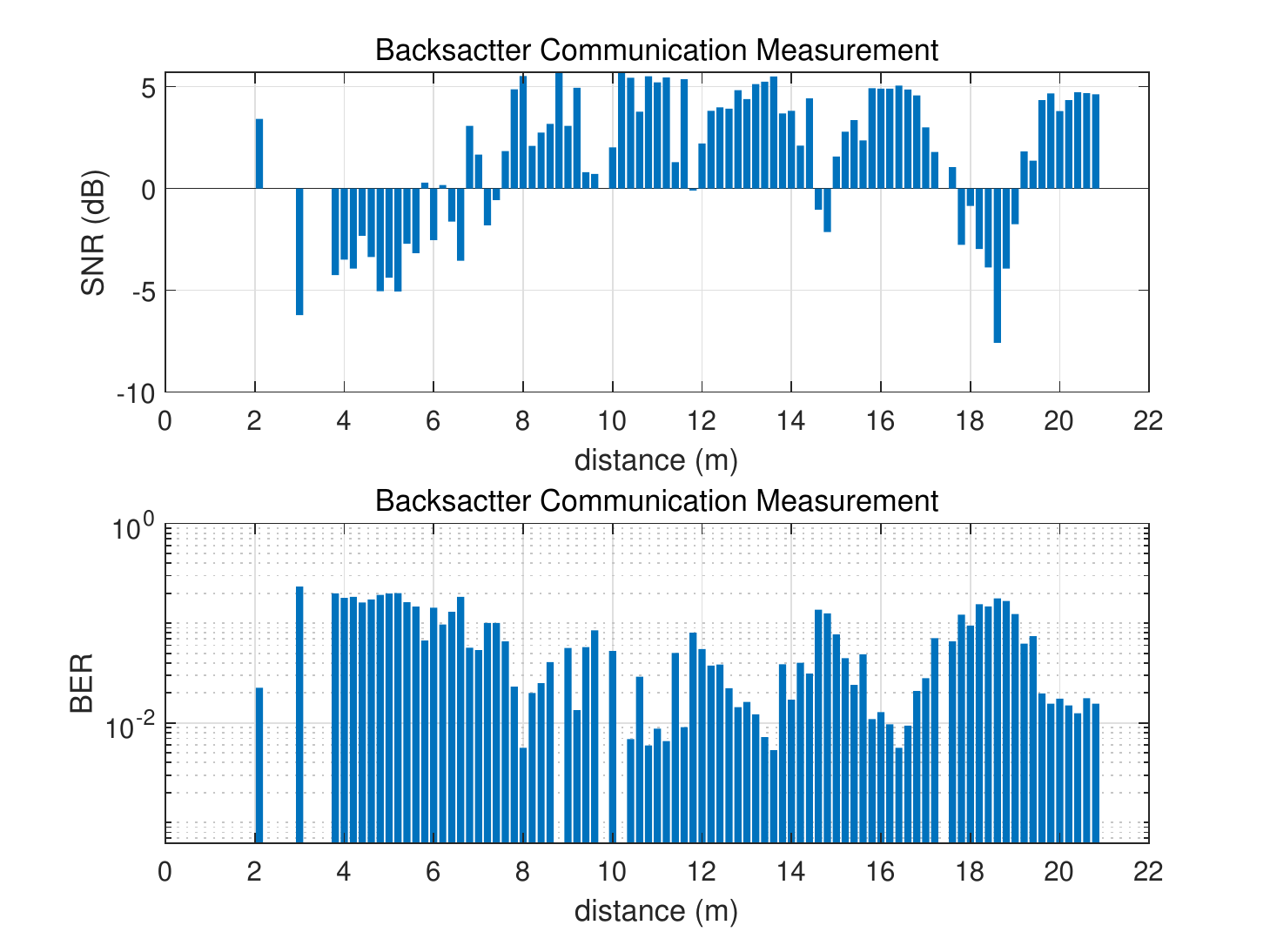} 
\caption{Wireless measured performance of AmBC in SNR and BER.}
\label{fig:validating_measure1204}
\vskip -12pt
\end{figure}

\begin{figure}[t!]
\centering
\includegraphics[width=0.85\columnwidth]{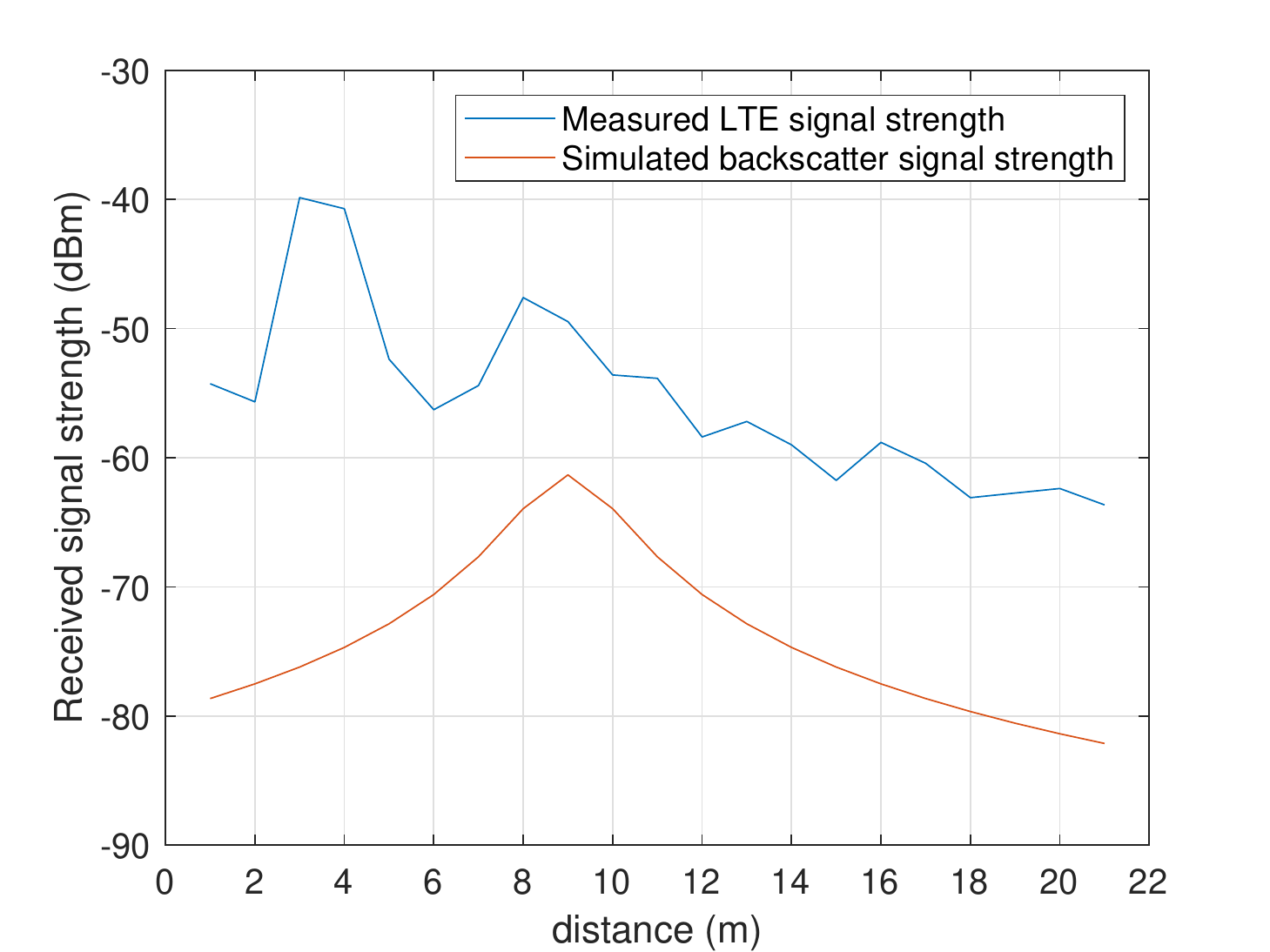}
\caption{Variation in the measured received LTE signal power and the simulated backscatter signal power at the 21 RX positions. The backscatter signal power is calculated using the FSPL with exponent of 2. \label{fig:powerInMeasPoints} }
\vskip -12pt
\end{figure}

Fig. \ref{fig:measurement_environment} presents the measurement environment in Maarintie 8 in Aalto University campus.
The transmitter antenna R\&S HK300 is located at the left lower corner in the Figure.
The BD node is in the middle of the corridor next to the measurement point 9 (at 9 m from the wall).
The receiver is USRP B205 unit and laptop running in house created C++ implementation of a LTE receiver, connected to UDP port with MATLAB implementation of AmBC signal detection.
The receiver implements real time decoding of received AmBC signal.

Differently from results reported in Fig. 5 in the previous work \cite{perviousPaper}, no external clock is distributed between transmitter, BD, and receiver.


In particular, the measurement system cannot detect weak signal which is lower than analog digital converter (ADC) in the environment.
In Fig. \ref{fig:powerInMeasPoints}, a simulation is given for received power of weak backscatter signal as a supplement. Simulation is based on the backscatter pathloss mode \cite{griffin2009complete} and completely ignores the impact of walls. It can be thus treated as an upper bound for the actual backcattered power.
\par
Data at some positions are lost, due to the high BER.
If BER is too high, a possible explanation is that backscatter frame is not correctly synchronized by three 7-barker bits header.
As subsection \ref{subsec:Synchronization} said, that data sequence is meaningless if packet is asynchronized.
In our measurement, if the BER is higher than one over three, then that backscatter data packet is omitted.
\par
In Fig. \ref{fig:validating_measure1204}, the SNR and BER are measured in the Fig. \ref{fig:measurement_environment} corridor, with step 0.2 m, from 2 m to 21 m.
Relationship between BER and SNR roughly satisfies the following relationship: low BER positions are usually under high SNR environments.
A sinusoidal tendency appears in SNR via distance.
That periodic phenomena in the tunnel corridor (6 m to 16 m) is distinct from other positions.
BER shows the same periodic regular pattern with SNR.
In the room 2536 (distance less than 6 m), SNR starts to deteriorate.
Then at the fork road corner (between 16 m and 18 m), SNR decreases steeply.
\par
Received ambient signal power at each measurement position is estimated based on received LTE signal power.
Step length of the LTE ambient signal power measurement is 1 m, which is blue line in Fig. \ref{fig:powerInMeasPoints}.
Backscatter signal power is calculated based on Friis transmission equation, which is red line in Fig. \ref{fig:powerInMeasPoints}.
We assume transmitter to BD and BD to each measurement position propagation paths are all line of sight.
FSPL module is applied in estimation of backscatter signal power, as Eq. (\ref{eq:fspl}).
But this FSPL module is higher than that of real backscatter signal power received on measurement positions.
In practice, the difference between ambient LTE signal power and backscatter signal power is even larger than that in Fig. \ref{fig:powerInMeasPoints}.
\par
Measured LTE signal power (blue line in Fig. \ref{fig:powerInMeasPoints}) for 8 m to 21 m approximately obeys FSPL.
Because BD is set at 9 m, simulated backscatter signal power (orange line in Fig. \ref{fig:powerInMeasPoints}) peaks at 9m.
It shows a typical FSPL pattern via distance.
Comparing the two plots, Fig. \ref{fig:validating_measure1204} and Fig. \ref{fig:powerInMeasPoints}, some noteworthy contrasts is given as following.
Around 6 m the LTE signal deteriorates.
That ambient signal attenuation also can be observed in Fig. \ref{fig:validating_measure1204}.
Around 6 m, SNR decreases dramatically with BER jumping to a high level.
A steep drop of LTE signal power from 1 m to 2 m is believed to be caused by a metal door between room 2045 and room 2536, which exactly between transmitter and measurement receiver.

\par

\section{Conclusions} \label{sec:Conclusions}
In this paper, we proposed a system that uses the LTE cell specific reference signals and channel estimator for receiving backscatter modulated signals. The BD utilized two square waves having different nominal frequencies to perform frequency shift modulation. The proposed receiver was validated using over the air measurements in an indoor environment. Based on our experimental results, we can conclude that the LTE channel estimator offers a great potential to be utilized for receiving backscattered signals in Ambient Internet of Things applications.

\section{Acknowledgements}
This work is in part supported by the European Project Hexa-X under grant 101015956 and Business Finland project eMTC under grant 8028/31/2022.


\bibliographystyle{IEEEtran}
\bibliography{References}
\vspace{11pt}
\vfill
\end{document}